\documentclass[sigconf,authorversion,nonacm]{acmart}
\AtBeginDocument{%
  }

\setcopyright{acmlicensed}
\copyrightyear{2025}
\acmYear{2025}
\acmDOI{XXXXXXX.XXXXXXX}
\acmConference[Conference acronym 'XX]{Make sure to enter the correct
  conference title from your rights confirmation email}{June 03--05,
  2018}{Woodstock, NY}
\acmISBN{978-1-4503-XXXX-X/2018/06}

\usepackage{comment}
\usepackage{adjustbox}
\usepackage{enumitem}
\setlist{nosep}

\usepackage[font=small]{caption}
\begin{document}

\title{Adapting Public Personas: A Multimodal Study of U.S. Legislators' Cross-Platform Social Media Strategies}

\author{Weihong Qi}
\email{wq3@iu.edu}
\affiliation{%
  \institution{Indiana University Bloomington}
  \city{Bloomington}
  \state{IN}
  \country{USA}
}

\author{Anushka Dave}
\email{daveanu@iu.edu}
\affiliation{%
  \institution{Indiana University Bloomington}
  \city{Bloomington}
  \state{IN}
  \country{USA}
}

\author{Chen Ling}
\email{ccling@iu.edu}
\affiliation{%
  \institution{Indiana University Bloomington}
  \city{Bloomington}
  \state{IN}
  \country{USA}
}


\begin{abstract}
 Current cross-platform social media analyses primarily focus on the textual features of posts, often lacking multimodal analysis due to past technical limitations. This study addresses this gap by examining how U.S. legislators in the 118th Congress strategically use social media platforms to adapt their public personas by emphasizing different topics and stances. Leveraging the Large Multimodal Models (LMMs) for fine-grained text and image analysis, we examine 540 legislators personal website and social media, including Facebook, X (Twitter), TikTok. We find that legislators tailor their topics and stances to project distinct public personas on different platforms. Democrats tend to prioritize TikTok, which has a younger user base, while Republicans are more likely to express stronger stances on established platforms such as Facebook and X (Twitter), which offer broader audience reach. Topic analysis reveals alignment with constituents' key concerns, while stances and polarization vary by platform and topic. Large-scale image analysis shows Republicans employing more formal visuals to project authority, whereas Democrats favor campaign-oriented imagery. These findings highlight the potential interplay between platform features, audience demographics, and partisan goals in shaping political communication. By providing insights into multimodal strategies, this study contributes to understanding the role of social media in modern political discourse and communications.
\end{abstract}

\begin{comment}
\begin{CCSXML}
<ccs2012>
 <concept>
  <concept_id>00000000.0000000.0000000</concept_id>
  <concept_desc>Do Not Use This Code, Generate the Correct Terms for Your Paper</concept_desc>
  <concept_significance>500</concept_significance>
 </concept>
 <concept>
  <concept_id>00000000.00000000.00000000</concept_id>
  <concept_desc>Do Not Use This Code, Generate the Correct Terms for Your Paper</concept_desc>
  <concept_significance>300</concept_significance>
 </concept>
 <concept>
  <concept_id>00000000.00000000.00000000</concept_id>
  <concept_desc>Do Not Use This Code, Generate the Correct Terms for Your Paper</concept_desc>
  <concept_significance>100</concept_significance>
 </concept>
 <concept>
  <concept_id>00000000.00000000.00000000</concept_id>
  <concept_desc>Do Not Use This Code, Generate the Correct Terms for Your Paper</concept_desc>
  <concept_significance>100</concept_significance>
 </concept>
</ccs2012>
\end{CCSXML}

\ccsdesc[500]{Do Not Use This Code~Generate the Correct Terms for Your Paper}
\ccsdesc[300]{Do Not Use This Code~Generate the Correct Terms for Your Paper}
\ccsdesc{Do Not Use This Code~Generate the Correct Terms for Your Paper}
\ccsdesc[100]{Do Not Use This Code~Generate the Correct Terms for Your Paper}

\end{comment}



\maketitle

\section{Introduction}

The use of social media platforms becomes increasingly important and widespread for politicians to communicate with citizens, campaign, and promote policies. With a significant portion of Americans turning to social media to stay informed about politics and political issues~\citep{Pew24}, American politicians placed considerable emphasis on social media outreach. The most active politicians post thousands of updates across various platforms~\citep{Statista23, straus2018social}. At the same time, survey report shows~\citep{Pew24} that Americans engage with different social media platforms for political discussions in distinct ways, leading to varied experiences. It is likely that many politicians respond to these differences in platform features and user demographics by employing different strategies tailored to each platform. For instance, TikTok users tend to be younger, while Facebook users are generally older~\citep{PewSocial}. Additionally, X (Twitter) allows for more lenient moderation, whereas Facebook and Instagram employ fact-checkers and enforce stricter policies on sensitive content.

With the growing importance of social media in political communication, extensive research has analyzed politicians' posts across multiple platforms. However, significant gaps persist in the current body of work. For example, some studies have comprehensively examined topics such as polarization~\citep{yarchi2021political}, campaigning~\citep{bossetta2023cross, niu2021teamtrees, panda2020topical}, and misinformation~\citep{lukito2020coordinating} over extended periods and with fine-grained analyses. However, there is a lack of research exploring why certain trends emerge on different platforms—an area that could be influenced by platform-specific features and user demographics. Additionally, while some studies have investigated multimodal content on social media, these efforts have been relatively small in scale and more general than fine-grained~\citep{farkas2021images}. Methodologically, advancements in large multimodal models (LMMs) offer powerful tools for large-scale, cross-platform, and multimodal analysis of political content. Yet, current research has not fully leveraged these tools to examine the strategies politicians use or the broader implications of these strategies.

In this research, we analyze the different strategies employed by 540 members of the 118th U.S. Congress across their official websites, Facebook, X (Twitter), and TikTok, utilizing large multimodal models (LMMs) for fine-grained multimodal analysis. Our study examines various aspects of social media promotion strategies, including positioning (stance), topics addressed, and post frequencies. Specifically, we uncover insights by linking these features to platform characteristics and user demographics, shedding light on how U.S. legislators perceive and engage with their constituents. Furthermore, we utilize LMMs such as {\tt Llama-3.1-8b}, {\tt LLaVA-1.5-7b}, and {\tt GPT-4o-mini} to perform multimodal analysis, examining image features such as photo type (content), presence of people, and visual formality. Additionally, we explore potential relationships among politicians' characteristics, text features, and their preferences for image usage in online promotions. Specifically, we investigate the following research questions:

\begin{itemize}
    \item \textbf{RQ1:} Do legislators portray themselves similarly or differently across various platforms?
% tailoring their messaging to the distinct audiences of each platform
    \item \textbf{RQ2:} Are there any partisan differences evident in topics and stances across different platforms?

    \item \textbf{RQ3:} Are there any associations between legislators' party affiliation, stances, and their preference for using multimodal content?
\end{itemize}

By creating a multimodal dataset of text and images, we examine the relationships between partisanship, stances, and image usage preferences in political promotions. Our findings indicate that legislators adapt their public personas based on the characteristics of each online platform. For instance, election-related content appears most frequently on TikTok, which caters to a younger audience. In contrast, platforms like Facebook and X (Twitter) are more commonly used for communicating constituent services, aligning with their broader and more diverse user bases. These patterns suggest that legislators strategically adjust their messaging to suit the features and audiences of each platform.

Our research contributes to the literature by analyzing the content and stances shared across various social media platforms to uncover the personas that legislators and their media teams aim to construct on each platform. This research also has significant implications for social media platform moderation policies and broader communication strategies.

%By comparing stances across platforms, we highlight how platform features and content moderation policies potentially shape varying levels of polarization. Furthermore, we provide insights into the alignment between the issues most important to constituents and the topics legislators address on their media platforms. 

\section{Related Work}
Social media platforms have rapidly become essential tools for political communication, enabling legislators and politicians to directly engage with voters and shape public discourse~\citep{panda2020topical, niu2021teamtrees, venancio2024unraveling, sapiezynski2024use}. Members of U.S. Congress increasingly rely on social media to connect with constituents, with near-universal adoption of established platforms like Facebook, X (Twitter), and YouTube~\citep{straus2018social}. In recent years, the proliferation of mobile cameras and the rapid dissemination of images on social media have made politicians more visible than ever~\citep{messaris2019digital}. This constant visibility pressures politicians to maintain a consistent and relatable public persona, both during and outside election cycles~\citep{baines2002political}. Through visual representation, politicians can make policies more tangible, providing symbolic and accessible depictions of abstract concepts~\citep{luntz2007words}. Despite the significance of multimodal content in political communication, there has been a notable lack of fine-grained, large-scale analyses in this area, largely due to technical limitations in earlier research. As recent research has highlighted the potential of using LMMs in computational social science~\citep{lyu2023gpt, ziems2024can}, these innovative models offer opportunities to expand the scope of data analysis.

Social media platforms have become essential in political communication because they allow legislators to reach broader audiences, using strategically tailored content and language to resonate with their base and enhance engagement. Carefully crafted word choices are critical for persuading audiences to align with a politician’s campaign. For instance, during the 2018 government shutdown, hashtags like \#TrumpShutdown (used by Democrats) and \#SchumerShutdown (used by Republicans) generated higher engagement on both X (Twitter) and Facebook~\citep{van2020congress}. Visual features also play a key role in shaping a candidate’s image as authentic and trustworthy, with physical attractiveness creating a halo effect that enhances perceptions of competence, trustworthiness, qualifications, and leadership ability~\citep{riggle1992bases, surawski2006effects}. Additionally, research highlights variations in language and framing across social media platforms and media outlets among contributors with different political leanings ~\citep{pan2023bias, pan2023understanding}. For example, conservatives tend to use slightly more power-related language, while liberals favor achievement-related language~\citep{jost2020language}. 
Building on previous research, our study conducts a multimodal content analysis to provide insights into the strategies legislators employ through the use of images and text.

Recent research investigates how legislators adapt their communication strategies across different social media platforms to enhance user engagement and strengthen connections with constituents. \cite{macdonald2023negative} explores how legislators use emotional appeals and negative sentiment on platforms like Facebook to foster partisan engagement and cue-taking among constituents, while \cite{russell2021tweeting} examines how U.S. senators utilize Twitter to craft rhetorical agendas that resonate with their constituents, highlighting the platform's role in shaping political representation. Additionally, cross-platform analyses reveal that politicians tailor their messaging to align with the unique affordances and user demographics of each platform~\citep{oliveira2022politicians, iazzetta2024cross, yarchi2021political}. For instance, research comparing Italian populist leaders' use of X (Twitter), Instagram, and Facebook demonstrates how message framing and content vary to maximize engagement on each platform~\citep{iazzetta2024cross}. These studies underscore the importance of understanding the various ways in which political communication is mediated by platform characteristics and audience expectations. Our multimodal content analysis contributes to this body of literature by examining how U.S. legislators employ text and imagery to construct tailored personas and engage with diverse constituencies across social media platforms.

Existing research also highlights the impact of multimodal political content on social media in shaping public opinions and decision-making~\citep{ausat2023role, neubaum2017monitoring, shah2024role}. For instance, the ``Echo Chambers'' perspective emphasizes the fragmented, highly customized, and niche-oriented nature of social media, suggesting that these platforms contribute to greater political polarization~\citep{hong2016political}. Given the significant role of social media in influencing public opinion and real-world decisions, understanding the strategies politicians use on these platforms becomes increasingly critical.

\section{Data Collection}

We collect data on 540 members of the 118th United States Congress (100 senators, 435 House representatives and five non-voting members) from various online platforms, including their official websites, X (Twitter), Facebook and TikTok. These platforms represent a spectrum of communication styles: the static, traditional format of official websites; the mature, widely adopted social media platforms Facebook and X (Twitter); and the emerging, video-centric platform TikTok. Together, they offer a diverse range of data, including text, images, and videos which enable the construction of a multimodal analytical framework. In addition to traditional textual data, we collect 103,925 thumbnail images from Facebook. The data collected for analysis spans the period from June 15 to November 15, including all messages posted by the legislators during the time. The selected time range captures the peak of the election campaign period, allowing the data to reflect legislators' cross-platform communication styles during the most active phase of the election year. By leveraging this multimodal dataset, researchers can uncover not only explicit information conveyed by politicians through text but also implicit relationships between text and visuals, shedding light on the social media strategies used to promote political messages.

For social media platforms, account retrieval is carried out through manual searches conducted directly on each platform. Specifically, we search the legislators' names according to the official website of the congress to identify their social media accounts. We use only publicly accessible information from the social media platform accounts of the legislators, ensuring that no private or sensitive data is collected. All data collection complies with the terms of service of the respective platforms. 

It is important to note that existing sources for identifying the official social media accounts of U.S. legislators, such as \cite{uscongress-git}, do not comprehensively capture their online activity. For example, we identified 41 legislators with active X (Twitter) accounts, including official ones, that are not documented in the cited source. Additionally, 13 legislators maintain non-official accounts that are more active than their official ones. To ensure more complete and accurate coverage, we therefore opted to manually collect account information rather than rely solely on existing datasets.

\subsection{Official Website} We use the Beautiful Soup library to scrape the official websites of legislators, which are listed on congressional webpages and designated as government sites. Each official website includes an ``Issues'' section that contains statements outlining the legislator's policy stances and the actions they have taken to support specific policies. To understand their stances on key issues, we collect all statements and texts under the ``Issues'' section of each official website, analyzing each issue separately. The dataset contains 6,038 statements from 288 legislators who provide valid statements about policies and issues on their websites. It is worth noting that text extraction using the Beautiful Soup library often includes noisy information, such as contact details. To address this, we use {\tt GPT-4o-mini} to clean the data by extracting only the relevant policy statements and removing extraneous content.

\subsection{Facebook} Next, we collect Facebook posts from the official pages of U.S. legislators for multimodal analysis, including both text and image data. The majority of legislators (515 out of 540) have active accounts with posts available from June 15 to November 15, 2024. In total, we gather 82,432 posts and 103,925 valid text entries and thumbnail images from 515 and 472 legislators, respectively. We use thumbnail images collected from Facebook, combined with text information extracted from the platform, for multimodal analysis. Facebook is chosen for this analysis due to its sufficient volume of images and accessible data.

\subsection{X (Twitter)} We collect user posts from legislators on X, covering the period from June 15 to November 15. It is noteworthy that a legislator may have multiple verified X (Twitter) accounts. In such cases, if one account has significantly more followers than the others, we select the account with more followers to construct the dataset for subsequent analysis. For example, Derrick Van Orden has two verified accounts, with follower counts of more than 489,000 and 6,411 as of April, 2025, respectively; thus, the account with significantly more followers is treated as the official account. We focus on the accounts with the most followers, as it is the most impactful channel for legislators to communicate with constituents. Additionally, accounts with fewer followers are excluded in cases of inactivity or a lack of recent posts. For example, Eric Swalwell’s less popular account did not post any tweets between November 11, 2024, and January 14, 2025, whereas his more popular account posted almost daily during that period, as well as before and after it. Including all accounts or uniformly focusing only on personal or campaign accounts could lead to an inaccurate representation of legislators' actual communication patterns on the platform. 

It is important to recognize that legislators often maintain multiple accounts—such as campaign, personal, and official accounts, each serving different purposes and employing distinct communication styles. In constructing our dataset, we focus on capturing communication patterns through the most impactful account, rather than differentiating strategies across all account types. For clarity and completeness, however, we also provide a comparison of official versus non-official accounts in the Results section, with detailed results presented in the Appendix.

\subsection{TikTok} We collect user information and posts from legislators on TikTok, also covering the same period from June 15 to November 15. Unlike X (Twitter), we find that only a small number of legislators, specifically 77, maintain a TikTok account. This suggests that having a TikTok account may reflect how legislators choose to engage with their constituents and whom they aim to reach. Given that TikTok's user base is younger compared to other mainstream social media platforms~\citep{PewSocial}, it indicates that these legislators may be focusing more on attracting young constituents. While we conduct a multimodal analysis, we do not include video data from TikTok, as its official API does not provide access to meaningful video content. Instead, we incorporate video descriptions and voice-to-text transcriptions from TikTok into our analysis.

Table~\ref{tab: des_stat} provides a summary of the overall data analyzed in this study. Notably, the number of posts across platforms is highly imbalanced, with Facebook and X (Twitter) containing significantly more posts compared to website issue statements and TikTok. This disparity highlights the dynamic and interactive nature of social media platforms, which facilitate greater engagement between legislators and constituents. It also suggests that legislators prioritize established platforms like Facebook and X (Twitter) over emerging ones like TikTok. Regarding the partisanship distribution of legislators' accounts, the data is relatively balanced between Democrats and Republicans on the website, Facebook, and X (Twitter) platforms. However, on TikTok, there is a significant disparity, with far more Democratic accounts than Republican ones, specifically, 64 Democrats compared to 11 Republicans. This imbalance suggests that Democrats are more actively using TikTok as a promotional and communication tool than Republicans.

\begin{table}
\centering
\adjustbox{max width=\linewidth}{
\begin{tabular}{{l}c*{4}{c}}
\toprule[1.1pt]
Platform   & \# of obs. & \# of legislators & \# of Democrats & \# of Republicans  \\
\midrule
Website & 6,038 & 288 & 149 & 137 \\
Facebook (text) & 82,432 & 515 & 248 & 261  \\
Facebook (image) & 103,925 & 472 & 232 & 231 \\
X (Twitter) & 115,192 & 533 & 256 & 270\\
TikTok & 2,886 & 77 & 64 & 11\\
\bottomrule[1.1pt]
\end{tabular}}
\caption{The descriptive statistics of the multimodal data collected from official website, X, Facebook and TikTok} 
\label{tab: des_stat} 
\end{table}

One common concern when using social media data for analysis is the highly skewed distribution of posts, where a small number of highly active users may disproportionately influence the results, potentially distorting broader patterns. To assess whether such ``super influencers'' exist in our dataset, we include in Table~\ref{tab:website_top10} - \ref{tab:tiktok_top10} in the Appendix the number of posts and their corresponding percentage of total posts for the top 10 most active legislators on each platform. On official websites, Facebook, and X (Twitter), no single legislator dominates the overall content; for example, the most active legislator, Jon Tester, contributes only 1.32\% of the total posts on official websites. The data for TikTok does reveal that a few Democrats contribute a disproportionately high number of posts. For instance, Jon Ossoff (10.50\%), Sherrod Brown (10.50\%), Collin Allred (7.03\%), and Jacky Rosen (6.62\%). However, Republican Katie Britt and J.D. Vance also accounts for a significant share of 7.10\% and 4.02\%, respectively. Given the imbalance in representation on the platform (64 Democrats compared to 11 Republicans), this distribution is not unexpected and will not distort the overall patterns identified in our analysis.

\section{Method}
\subsection{Data Preprocessing}
To structure and clean the data for further analysis, we utilize Large Multimodal Models (LMMs), including both open-source and commercial models such as {\tt Llama-3.1-8b}, {\tt LLaVA-1.5-7b}, and {\tt GPT-4o-mini}, for preprocessing and labeling. Notably, we observe significant imbalances in text length across our dataset—for example, official website issue statements are much longer than social media posts. To address this, we select models based on the characteristics of the data and cost considerations. Specifically, we use the {\tt Llama-3.1-8b} model to label social media text data due to its cost efficiency and suitability for shorter texts~\citep{touvron2023llama}. For processing longer text from official websites, we utilize the commercial model {\tt GPT-4o-mini}, which performed better in handling lengthy text and delivered more valid results in our pilot tests. Finally, we employ the {\tt LLaVA-1.5-7b} model for processing image features, as it is both cost-effective and demonstrates state-of-the-art performance in various image analysis tasks~\citep{liu2023llava, liu2023improvedllava}.

We begin by using {\tt GPT-4o-mini} to clean the data collected from official websites, extracting statements from text that contains significant noise. {\tt GPT-4o} is an autoregressive omni-model developed by OpenAI, and {\tt GPT-4o-mini} is a smaller, cost-efficient version of the model~\citep{hurst2024gpt, gpt4omini}. Since {\tt GPT-4o-mini} demonstrates comparable performance to {\tt GPT-4o} on our dataset, we select this model for cleaning the text data. We provide the specific prompt used, along with examples of raw data and extracted statements processed for analysis.

\subsection{Data Labeling}
In our research, we focus on analyzing the topics and stances expressed in the text data. These two features, representing the qualitative and quantitative aspects of posts, are central to political communication research~\citep{stieglitz2013social}. To minimize subjective bias in data labeling and assess labeling performance, we randomly sample observations from each social media platform dataset and the official website dataset, and compare the results between two human annotators as well as between human annotations and model-generated annotations across all text labeling tasks. We begin by calculating the inter-annotator agreement score as a benchmark for evaluating human-model agreement. Both annotators are the authors of this paper and have academic backgrounds in political science and international studies, equipping them to provide informed and contextually grounded annotations. To assess inter-annotator agreement, each annotator independently labeled 300 observations from the social media datasets and 150 from the website dataset, with 150 and 75 overlapping observations, respectively, used to measure consistency. After the process, the human annotators coordinated to revise the labels for agreement on the overlapping samples. These human-labeled datasets are then used to evaluate model accuracy, defined as the level of agreement between model-generated labels and human annotations. Specifically, both inter-annotator agreement and human-model agreement are calculated using the following agreement score:

\begin{equation}
Agreement = \frac{\sum_i^{N} \mathbb{I}_i(A_1 =A_2)}{N}
\end{equation}

In this formula, $A_1$ and $A_2$ represent the labels assigned by the two annotators, and $\mathbb{I}_i$ is an indicator function that equals 1 if the two values are considered equivalent, and 0 otherwise. $N$ denotes the total number of observations. Note that we do not use other common metrics, such as Cohen's Kappa, as they are not ideal for evaluating agreement in labeling tasks involving ranked or ordinal labels, as used in our research. Instead, we rely on an agreement score that is validated by~\cite{argyle2023out}.

\subsection{Topic Labeling} 
We define the topics for labeling based on the most salient issues of concern to American citizens during the election year~\citep{PewIssue}. Additionally, since social media platforms serve as key channels for legislators to communicate about their day-to-day services, the topics also include constituent services. The categories are constituent service, economy, health care, foreign policy, immigration, gun control, abortion, racial and ethnic inequality, climate change, election, and other. We use {\tt GPT-4o-mini} to label topics in official website statements and {\tt Llama-3.1-8b} to label topics in data from Facebook, X (Twitter), and TikTok. Notably, we assign multi-class labels in our research, because social media posts are typically brief and tend to focus on a single topic per post. Similarly, the statements from official websites are scraped from the "Issues" section, where each page generally addresses one specific topic. Therefore, a multi-class labeling scheme is more appropriate for our dataset and research objectives. Table~\ref{tab: topic_eval} presents the model performance evaluation for topic labeling. The specific prompt used in topic labeling is provided in the Appendix. Overall, the model's accuracies are comparable to the human agreement levels, which ranges from 0.6 to 0.8. This is considered satisfactory for an eleven-category labeling task, as random guessing would result in an accuracy of approximately 0.09.

\begin{table}[h!]
\centering
\adjustbox{max width=\linewidth}{
\begin{tabular}{{l}c*{2}{c}}
\toprule[1.1pt]
Platform   & Human Agreement & Model Accuracy &  \\
\midrule
Website & 0.6471 & 0.6423 \\
Facebook & 0.8523 & 0.7933  \\
X (Twitter) & 0.7533 & 0.7867 \\
TikTok & 0.7133 & 0.6633 \\
\bottomrule[1.1pt]
\end{tabular}}
\caption{Evaluation of Model Performance in Topic Labeling} 
\label{tab: topic_eval} 
\end{table}

\subsubsection{Stance Labeling} 
To capture the varying strengths of opinions expressed in the text, we assign numerical values ranging from -3 to 3 to represent stances, where -3 indicates the most left-leaning stance, 3 represents the most right-leaning stance, and 0 is neutral. Note that we use a seven-point scale [-3, -2, -1, 0, 1, 2, 3] for stance labeling, rather than a simpler scale such as [-1, 0, 1] or [-2, -1, 0, 1, 2], to capture not only the direction but also the strength of legislators’ political leanings. We acknowledge that this finer-grained scale introduces a trade-off between granularity and labeling accuracy, which may contribute to relatively lower performance in LMM-based labeling. Therefore, we evaluate model performance against human agreement benchmarks.

It's important to note that many LMMs tend to avoid extreme responses in practice, which may introduce bias in stance labeling. To address this, we follow the approach outlined in \cite{zhang2024electionsim}, prompting the LMMs with a scale of -10 to 10 and applying the following mapping for evaluations and data analysis:

\begin{equation}
f(x) = 
\left\{
\begin{array}{ll}
0, & \text{if } -1 \leq x \leq 1, \\
-1, & \text{if } -4 \leq x \leq -2, \\
-2, & \text{if } -7 \leq x \leq -5, \\
-3, & \text{if } -10 \leq x \leq -8, \\
1, & \text{if } 2 \leq x \leq 4, \\
2, & \text{if } 5 \leq x \leq 7, \\
3, & \text{if } 8 \leq x \leq 10
\end{array}
\right.
\end{equation}

In this equation, $x$ represents the stance score given by the LMMs, and $f(x)$ denotes the final labeled score. 

Table~\ref{tab: stance_eval} presents the human agreement and model evaluation for stance labeling. Notably, the agreement and model performance for stance labeling are lower compared to topic labeling, primarily due to the greater subjectivity involved in stance assessment. Model accuracy is particularly low for website stance labeling, likely because of the longer text length. However, as shown in the Appendix, most mislabeled data are close to the ground truth (e.g., the model labels -3 instead of -2 or -1 instead of 0). Despite this, the agreement between human annotations and model accuracy remains comparable and significantly higher than random guessing for a seven-category labeling task. Therefore, we consider the labeling performance satisfactory, and the minor mislabeling is unlikely to affect the main conclusions of our research. 

\begin{table}[h!]
\centering
\adjustbox{max width=\linewidth}{
\begin{tabular}{{l}c*{2}{c}}
\toprule[1.1pt]
Platform   & Human Agreement & Model Accuracy &  \\
\midrule
Website & 0.6176 & 0.5255 \\
Facebook & 0.7584 & 0.7407  \\
X (Twitter) & 0.6933 & 0.7267\\
TikTok & 0.6467 & 0.6309 \\
\bottomrule[1.1pt]
\end{tabular}}
\caption{Evaluation of Model Performance in Stance Labeling} 
\label{tab: stance_eval} 
\end{table}

\subsubsection{Image Labeling} 
We build on the tradition of political image analysis by categorizing image features into three main groups: content, presence of people, and visual formality~\citep{farkas2021images}. The content category includes features such as campaign, official, personal, and policy-related images. The presence of people category identifies whether images feature groups, individuals, or no people. Finally, the visual formality category classifies images as formal or informal. For the content category, we directly use LMMs for labeling. However, we find that while {\tt LLaVA-1.5-7b}'s performance in identifying the presence of people is limited, its descriptions of images are accurate\footnote{Detailed prompts and examples of image descriptions are provided in the Appendix}. This limitation is consistent with findings in the existing literature~\citep{dong2024benchmarking}. To address this, and drawing on lightweight cascade methods that avoid additional detection fine-tuning while leveraging the reasoning capabilities of LMMs~\citep{ye2019cap2det, rotstein2024fusecap}, we prompt {\tt LLaVA-1.5-7b} to generate one-sentence image descriptions and then use {\tt GPT-4o-mini} to extract information about the presence of people. Finally, for images that include people, we use {\tt GPT-4o-mini} to detect keywords ``suit'' and ``suits'' in the image descriptions, allowing us to assess the level of visual formality.

\begin{table}[h!]
\centering
\adjustbox{max width= \linewidth}{
\begin{tabular}{{l}{l}c*{2}{c}}
\toprule[1.1pt]
Category & Feature   & Model Accuracy  \\
\midrule

   Content & Campaign, Official, Personal, Policy  & 0.8200  \\

 Presence of People   & Group, Individual, No People & 0.8900 \\

Visual Formality & Formal, Informal &  0.7950  \\

\bottomrule[1.1pt]
\end{tabular}}
\caption{Evaluation of Model Performance in Image Labeling} 
\label{tab: img_eval} 
\end{table}

To evaluate the capability of LMMs in labeling the image features of interest, we randomly sample 200 images from our dataset. These images are annotated by one of the authors, who has advanced academic training in political science and data science. Table~\ref{tab: img_eval} presents the model evaluation results for image labeling. Unlike the stance labeling task, model accuracy is substantially higher for image labeling, likely due to the relatively simpler and clearer classification criteria. With accuracy ranging from 0.8 to 0.9, which is well above the level of random guessing, we consider the model's performance satisfactory for generating image labels for further analysis.

\subsection{Statistical Analysis}
In the image analysis, we primarily use correlation coefficients and p-values to examine the relationship between stances and the image features employed. To highlight partisan differences, we report t-test results and corresponding p-values for the use of image features by the two major political parties in the United States. The formula used for the t-test is as follows:

\begin{equation}
t = \frac{\bar{X}_D - \bar{X}_R}{\sqrt{\frac{s_D^2}{n_D} + \frac{s_R^2}{n_R}}}
\end{equation}

\begin{figure*}[h!]
    \centering
    \includegraphics[width=.9\textwidth]{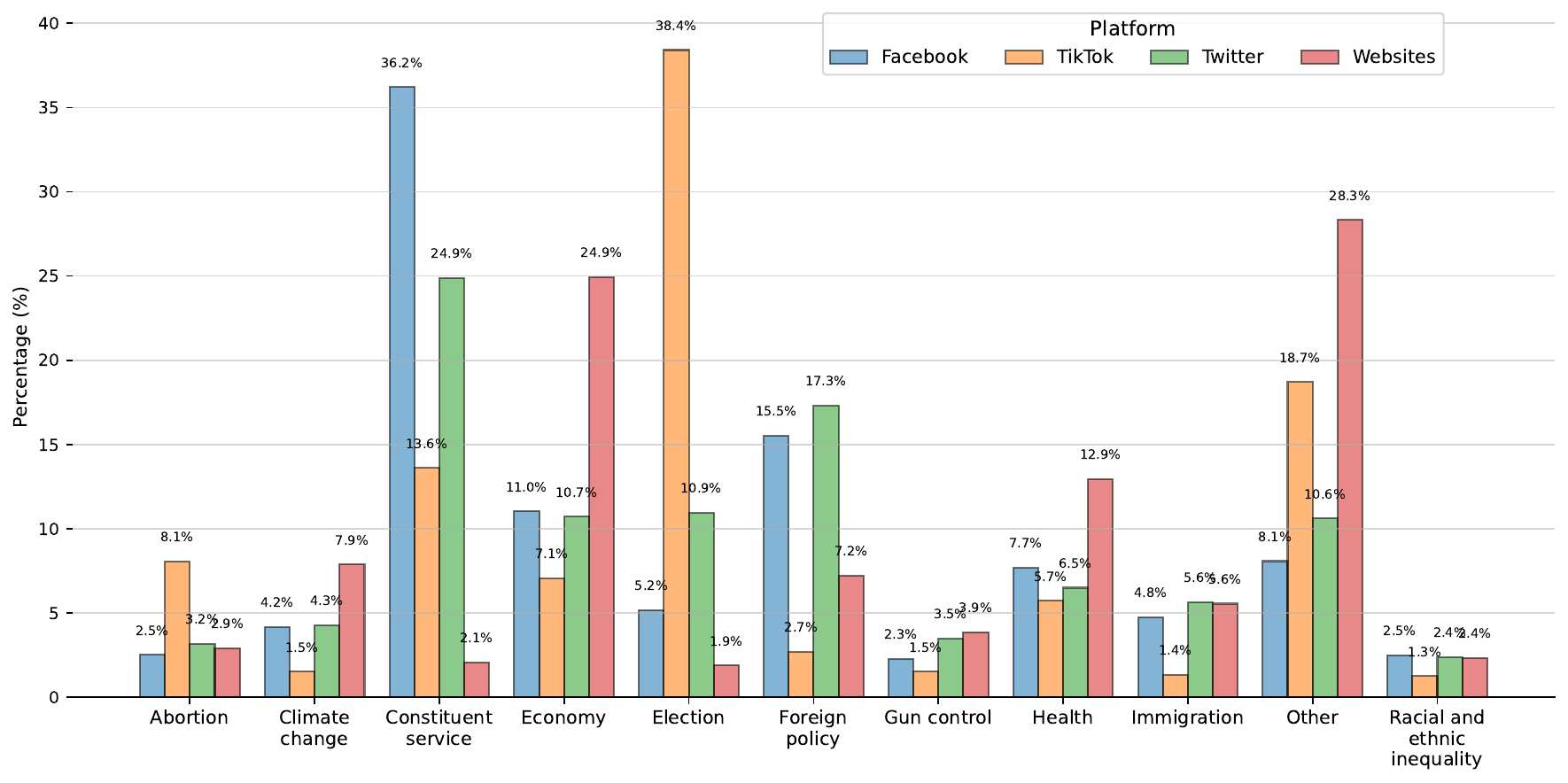}
    \caption{Topic Percentage Distribution by Platform.}
    \label{fig: topic_d}
\end{figure*}

Here, group $D$ represents Democrats, and group $R$ denotes Republicans. $\bar{X}$ represents the sample mean, $s$ denotes the variance of the group, and $n$ is the sample size of the group. A positive t-statistic indicates that the use of a particular feature is higher among Democrats than Republicans, while a negative t-statistic suggests the opposite.

\section{Results}

\subsection{Official vs. Non-official Accounts}

As mentioned in the Data Collection section, we are aware that legislators may have multiple social media accounts, especially on the X (Twitter) platform. The official accounts, which are sponsored by the Congress, is not allowed to be used for campaign activities such as fund raising. Although our analysis focus on the most impactful communication channels of the legislators, to avoid the bias when mixing the accounts in the analysis, we also present the official vs. non-official account post comparison with X (Twitter) accounts before we head into more detailed analysis. 

First, as shown in Figure~\ref{fig: balance_party} in the Appendix, the distribution of partisanship across the most active accounts—whether official or non-official—is relatively balanced: 86\% of Democrats and 84\% of Republicans have their official account as the most active. Figure~\ref{fig: balance_topic} in the Appendix further illustrates that official accounts post more frequently about constituent services, whereas non-official accounts focus more on elections, which aligns with existing regulations. Since constituent service posts are excluded from our subsequent analysis, this difference does not affect the main conclusions presented in the following sections.

We then examine whether official and non-official accounts differ in their expressed stances. As shown in Figure~\ref{fig: balance_stance}, both types of accounts display similar stance patterns across topics. Overall, the results indicate that official and non-official accounts are relatively balanced across parties and that their content does not differ substantially in the topics under study. Thus, it is reasonable to analyze them together in the subsequent sections.

\subsection{Topic Distribution}

In this subsection, we analyze the topic distribution across platforms and connect it to platform-specific features to uncover legislators' cross-platform strategies. Figure~\ref{fig: topic_d} illustrates the distribution of topics across different platforms. The most frequently addressed topics vary significantly by platform. For example, election-related content is predominantly posted on TikTok, while Facebook and X (Twitter) serve as primary platforms for communicating constituent services. Given that TikTok's user base skews younger compared to the other two platforms, this suggests that legislators view TikTok as a key platform for mobilizing younger voters, while using Facebook and X (Twitter) to demonstrate their commitment to day-to-day services for older constituents. In addition to the categories identified, the ``Other'' topic primarily includes content related to constituent services and various policy endorsements. On official websites, this category often features information about veteran services. On social media platforms, it encompasses posts celebrating events such as Independence Day and Pride Month, as well as personal glimpses into legislators’ lives, such as spending time with their families.

According to \cite{PewIssue}, American citizens identified the economy, health care, and foreign policy as the most important topics during the election year. Legislators addressed these concerns by prioritizing content on these topics, making them the top issues discussed on their official websites. A similar trend is evident on Facebook and X (Twitter), reflecting that U.S. legislators are attuned to their constituents' priorities and actively engage with them on various platforms.

\subsection{Stance Analysis}

\begin{figure}
    \centering
    \includegraphics[width= \linewidth]{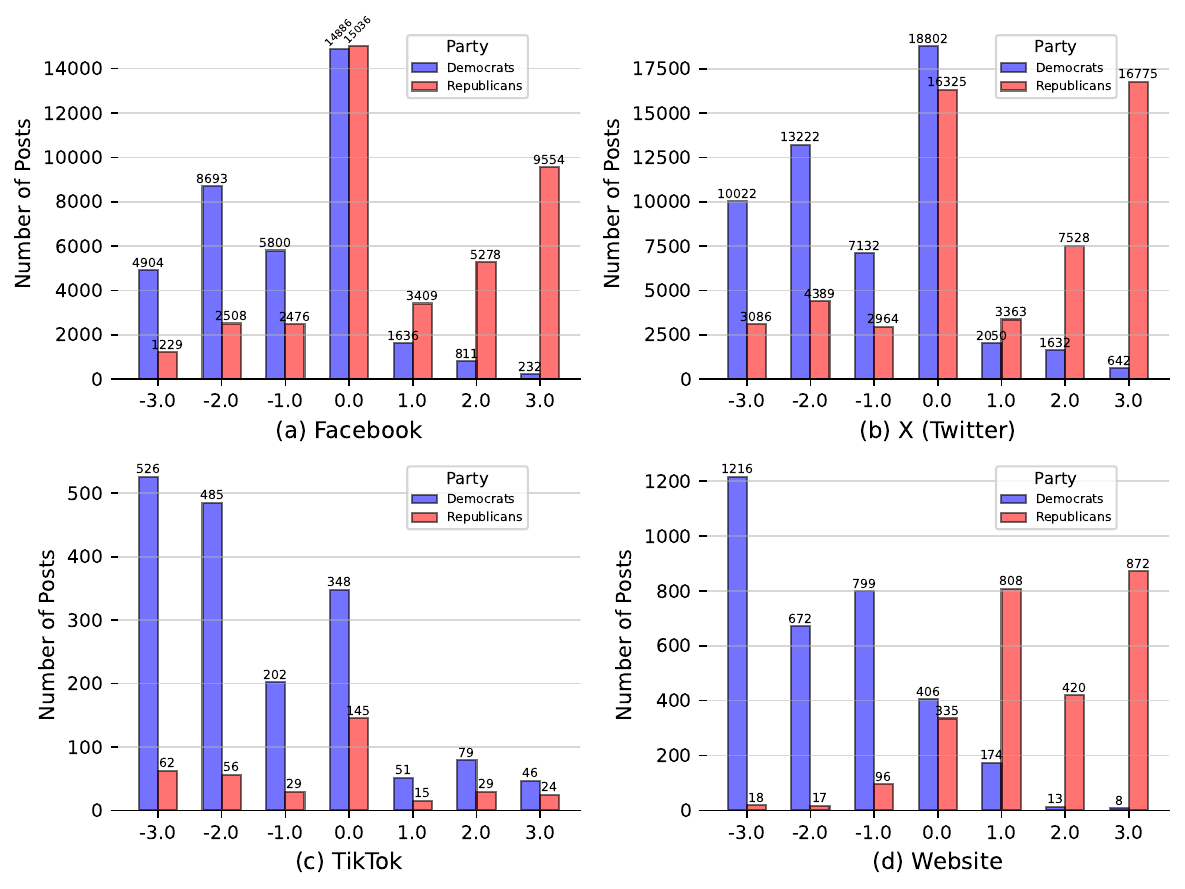}
    \caption{Stance Score Distribution by Platform and Party Affiliation}
    \label{fig: stance_party}
\end{figure}

\begin{figure}[h!]
    \centering
    \includegraphics[width= \linewidth]{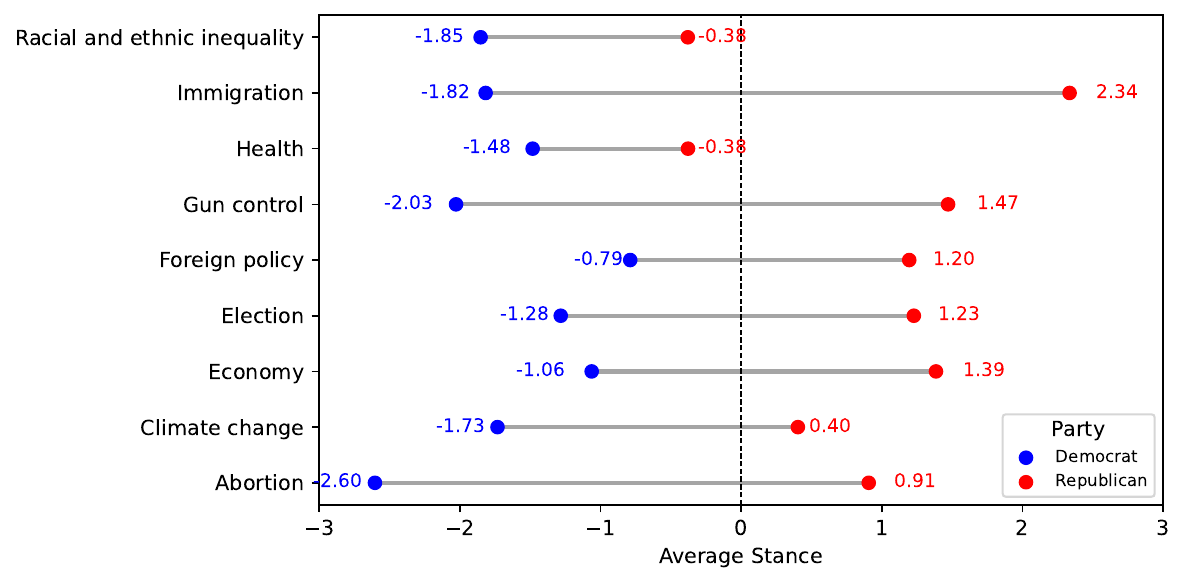}
    \caption{Stance Divergence Scores Across Topics by Partisanship}
    \label{fig: div_party}
\end{figure}
Using the stance scores labeled by LMMs, we analyze stances by partisanship across different platforms. Figure~\ref{fig: stance_party} presents the stance distribution for each platform. Notably, the two parties display varying levels of opinion strength depending on the platform. For instance, Republicans express stronger opinions on Facebook and X (Twitter) compared to Democrats, whereas Democrats exhibit stronger opinions on TikTok and official websites.  Additionally, Democrats post significantly more content than Republicans on TikTok, reflecting not only their greater presence on the platform but also a strategic emphasis on engaging and mobilizing younger voters. The variation in stances across platforms, combined with differences in topic emphasis, suggests that U.S. legislators strategically tailor their positions and content on each platform to maximize engagement with different audiences.
%This suggests that legislators tailor their content to maximize engagement: Republicans may perceive that more extreme content drives higher engagement on established platforms like Facebook and X, which have broader user bases. Conversely, Democrats appear to believe that more extreme content is more effective on emerging platforms like TikTok, which have a younger and more niche audience. 

Figure~\ref{fig: div_party} illustrates the levels of polarization across topics by partisanship, represented by the average stances of legislators from the two major U.S. political parties. Among all topics, racial and ethnic inequality and health care exhibit the lowest levels of polarization. While Democrats hold moderately left-leaning stances on these issues, Republicans' stances are closer to neutral. Conversely, immigration emerges as the most polarized topic, reflecting its prominence as a key issue in the 2024 election~\citep{PewIssue}.

\subsection{Image Analysis}

Table~\ref{tab:img_corr} summarizes the correlation between politicians' average stances and their patterns of image use on social media. By combining LMM-labeled image features with statistical analysis, we are able to uncover the correlations. Overall, there is no significant correlation between stance and the content categories, as none of the four content features demonstrate a notable relationship. However, official content shows a slight positive correlation with conservative legislators. A similar trend is observed in the presence of people in images, with no significant relationship between stance and who appears in the images. Finally, conservative legislators exhibit a higher degree of visual formality in their images compared to liberal legislators.

The t-test results in Table~\ref{tab:img_tt} highlight partisan differences between Democrats and Republicans in their use of images. The findings suggest that Democrats are more likely to post campaign-related images, while Republicans show a slight preference for official and more formal images. Combined with the correlations between stance and image features, these results indicate that Republican legislators tend to project greater visual formality compared to Democrats, who focus more on campaign imagery. Since visual formality conveys a more authoritative image, this suggests that Republicans aim to portray themselves as more authoritative, whereas Democrats prioritize visual appeal to mobilize voters.

\begin{table}[h!]
\centering
\adjustbox{max width=\linewidth}{
\begin{tabular}{{l}{l}c*{2}{c}}
\toprule[1.1pt]
Category & Feature   & Correlation & P-value &  \\
\midrule
Content & Campaign & -0.0602 &  0.1915 \\
& Official & 0.0892 & 0.0527 \\
& Personal & -0.0338 & 0.4641 \\
& Policy & -0.0315 & 0.4943 \\
\midrule
Presence of People & Group & 0.0075 & 0.8704 \\
                & Individual & 0.0531 & 0.2495 \\
                & No People & -0.0359 & 0.4361 \\
\midrule
Visual Formality & Formal & 0.1090 & 0.0179 \\
                & Informal & -0.0269 &  0.5606 \\
\bottomrule[1.1pt]
\end{tabular}}
\caption{Correlation Between Image Feature Counts and Legislators' Average Stances} 
\label{tab:img_corr} 
\end{table}

\begin{table}[h!]
\centering
\adjustbox{max width=\linewidth}{
\begin{tabular}{{l}{l}c*{2}{c}}
\toprule[1.1pt]
Category & Feature   & T-Statistic & P-Value &  \\
\midrule
Content & Campaign & 2.0492 & 0.0410 \\
        & Official & -1.8039 & 0.0719 \\
        & Personal &  0.8519 &  0.3947 \\
        & Policy & 0.2177 &  0.8277 \\
\midrule
Presence of People & Group & 0.0757 &  0.9397\\
            & Individual & -0.9722 &  0.3314\\
             & No People & 1.3643 &  0.1731 \\
\midrule
Visual Formality & Formal & -1.9344 &  0.0537 \\
                & Informal & 0.8411 & 0.4007 \\
\bottomrule[1.1pt]
\end{tabular}}
\caption{T-Test Results for Image Feature Usage Between Democratic and Republican Legislators} 
\label{tab:img_tt} 
\end{table}

\section{Conclusions}
Our study examines how U.S. legislators in the 118th Congress use social media platforms to engage with constituents, focusing on variations in topics, stances, and image usage across platforms. Utilizing large multimodal models (LMMs) such as {\tt GPT-4o-mini}, {\tt LLaVA-1.5-7b}, and {\tt Llama-3.1-8b}, we perform a fine-grained analysis of textual and visual content to uncover key insights into political communication strategies. By employing LMMs for text and image labeling, we effectively extract essential information from legislators' social media posts, enabling a deeper understanding of their platform-specific approaches.

We find that legislators tailor their topics and stances across platforms to project different public personas. For example, election-related content is most prevalent on TikTok, a platform with a younger demographic. Conversely, Facebook and X (Twitter) are used more frequently for constituent service communication, appealing to their broader and more diverse user bases. These platform-specific strategies reflect an awareness among legislators of the need to adapt their communication style on different platforms.

%, suggesting that legislators view it as a critical space to engage and mobilize young voters

In terms of stances, we observe significant variations by platform and partisanship. Republicans express stronger opinions on Facebook and X (Twitter), while Democrats exhibit stronger opinions on TikTok and their official websites. This suggests that Republicans may prioritize established platforms with larger audiences to amplify their messaging, while Democrats emphasize emerging platforms like TikTok to reach and engage younger voters. Additionally, Democrats post significantly more content on TikTok than Republicans, further underscoring their strategic focus on this demographic.

Our analysis of topic distributions reveals that the economy, health care, and foreign policy are consistently the most discussed issues across platforms, aligning with the primary concerns of American citizens during the election year~\citep{PewIssue}. This alignment suggests that legislators are aware of their constituents' priorities and actively address these issues in their messaging. However, the levels of polarization vary by topic, with immigration being the most polarized and health care and racial inequality exhibiting the least polarization. These findings highlight the nuanced ways in which partisanship shapes political discourse across topics.

In examining image use, we find clear partisan differences. Republicans are more likely to use images with higher visual formality, such as those featuring suits and official settings, to project authority and competence. Democrats, on the other hand, are more inclined to post campaign-related images, emphasizing visual appeal to mobilize voters. These strategies reflect differing approaches to leveraging visual content: Republicans focus on establishing an authoritative persona, while Democrats prioritize engagement through relatable and visually dynamic content.

Overall, our study demonstrates how legislators strategically adapt their personas to align with platform features, audience demographics, and partisan goals. By combining text and image analysis, we provide a comprehensive understanding of the interplay between political stances, topics, and visual content in shaping modern political communication. These findings contribute to the broader understanding of how social media platforms mediate political discourse and offer valuable insights into the implications of platform-specific communication strategies on public opinion and voter engagement.

\section{Discussions}

Our study, while comprehensive, acknowledges several limitations, particularly its focus on U.S. legislators' social media activity. The unique political, cultural, and institutional contexts of the United States may influence both the data patterns observed and the strategies employed by legislators. As such, the methods and conclusions of this study should be carefully adapted when applied to other countries, where differences in political systems, social media usage patterns, and audience expectations could significantly alter the findings. Our research is also limited by its focus on the descriptive aspects of U.S. legislators’ online communication, rather than offering causal evidence of strategic intent or user engagement across platforms. Future studies can address this gap by incorporating causal analyses to better understand the underlying strategies.

Additionally, our analysis is limited by the availability of data, particularly the lack of access to video content on platforms like TikTok, which prevents a deeper exploration of video-based communication strategies. Future research could expand the scope to include politicians from diverse regions and utilize advanced methods to analyze more complex multimodal data, such as videos and audio. These extensions would provide a more comprehensive understanding of cross-platform political communication globally. Our focus on the most popular X (Twitter) accounts allows us to examine communication patterns through the most impactful channels but limits our ability to differentiate strategies across various types of accounts, such as personal versus campaign profiles. Additionally, the temporal nature of the dataset restricts the study to the most active phase of the election year, and caution should be exercised when generalizing the findings to other political periods as noted in existing research~\citep{munger2019limited}.

To maintain the focus of this paper on the relationship between platform features, textual content, and image use patterns, we limit our analysis to the most common image features rather than conducting a detailed exploration of diverse image characteristics. Additionally, while state-of-the-art models like {\tt LLaVA-1.5-7b} and {\tt GPT-4o-mini} excel in many tasks, they exhibit notable limitations in identifying political features within images. These constraints further limit the depth of our image analysis. Future work could address these challenges by leveraging more advanced or specialized models to conduct a more comprehensive exploration of political image features.

%\begin{acks}
%To Robert, for the bagels and explaining CMYK and color spaces.
%\end{acks}

\bibliographystyle{ACM-Reference-Format}
\bibliography{sample-base}

\appendix

\section{Appendix}

\subsection{Data Preprocessing and Labeling Prompts}
\label{sec:prompt}

We use the following prompt in {\tt GPT-4o-mini} model to clean the text we collect from official websites of legislators. Specifically, the prompt instructs {\tt GPT-4o-mini} to retain the main policy stance statements of the legislators while removing irrelevant information such as contact details and excessive punctuation. This cleaned data streamlines the labeling process for human annotators and enhances the accuracy of text annotation using LLMs.

\textit{\textbf{Prompt 1.1:}} Please clean the text scraped from a U.S. legislator's website. If you determine that the text contains an official policy statement, proceed accordingly; otherwise, respond with ``no information'':\{text\}.
The following example shows how the essential information is extracted from the noisy text data.

\textit{\textbf{Example 1.1}}: 

Raw data: ['Valley Fever Task Force Website', 'In 2013, Congressman David Schweikert and Majority Leader Kevin McCarthy teamed up to co-found the Congressional Valley Fever Task Force, bringing much needed awareness to the disease of coccidioidomycosis.\  The disease, more commonly known as Valley Fever, is prominent in Maricopa county as well as the rest of Arizona and southern California.', 'The goal of the Valley Fever Task Force is to share information with stakeholders in the medical and scientific fields to foster new advancements in prevention and treatment as well as work with community organizations to help educate individuals on the disease.\  As Congressman Schweikert stated,\ “This task force is a much needed step toward raising awareness for this terrible disease and someday soon finding a cure. Valley Fever has silently affected entire communities in the southwest including our family, friends, and even beloved pets.\ I am hopeful that this working group will bring awareness, reduce the risk of misdiagnoses, and bring about a cure within the decade.”', 'In the 113th Congress, Congressman Schweikert successfully led the effort to have coccidioides spp., the pathogens that cause Valley Fever, listed as qualifying pathogens under the GAIN Act of 2011.\  This effort granted Valley Fever the title of “orphan disease” with the Food and Drug Administration (FDA), meaning any treatments or future cures will be given priority and fast tracked through the often arduous FDA drug approval process.', 'What is Valley Fever?', 'Valley Fever is caused by the fungal spore coccidioides spp. endemic to arid and semi-arid geographical areas. Because these spores are carried by soil, any disruption to the ground creates a heightened risk of infection. While some individuals breathe in the spore with no repercussions, others fall ill from Valley Fever.\  Those affected most by Valley Fever are immunosuppressed patients; elderly individuals; pregnant females; and minority populations of African, Filipino, and Native American descent.\ Of the more than 150,000 individuals infected annually, roughly 50,000 warrant medical attention.\  Of those, nearly 600 cases have the infection spread from their lungs to other parts of their body. Ultimately, about 160 cases result in death. Reported cases of Valley Fever in the southwestern states of Arizona, California, Texas and New Mexico have skyrocketed in recent years; the Centers for Disease Control and Prevention (CDC) statistics show that reported cases of Valley Fever have risen nine-fold since 1998.\  Arizona alone accounts for about 65\% of all reported cases that occur in the United States.', 'Today, patients diagnosed with Valley Fever are prescribed an antifungal therapy that has a 30\%-60\% failure rate. The current recommendation of the Infectious Disease Society of America is to continue treatment for life. There is no cure.', 'Valley Fever severely affects many who work outdoors, farmers, and those in the construction industry. Every branch of the United States’ military has bases in the endemic regions, putting our military men and women at risk.', 'Press Release', 'December 12, 2023', 'WASHINGTON, D.C.\ —\ Today,\ U.S. Representative David Schweikert (AZ-01) and former Speaker Kevin McCarthy (CA-20), co-chairmen of the\ Congressional Valley Fever Task Force, and…', 'Press Release', 'September 13, 2023', 'WASHINGTON, D.C.\ —\ Yesterday, Congressman David Schweikert (AZ-01) and Speaker Kevin McCarthy (CA-20), co-chairmen of the Congressional Valley Fever Task Force, led…', 'Press Release', 'October 14, 2021', 'WASHINGTON, D.C. – Today,\ Congressional Valley Fever Task Force Co-Chair Congressman David Schweikert (AZ-06) alongside Task Force Co-Chair Congressman Kevin McCarthy…', 'In The News', 'September 01, 2021', 'TUCSON, Ariz. (KOLD News 13) – The coronavirus pandemic is intersecting with another respiratory illness endemic to Arizona: Valley Fever….', 'Office Locations', '  Washington, DC Office 460 Cannon House Office Building Washington, DC 20515   Phone: (202) 225-2190   Fax: (202) 225-0096 ', '  Scottsdale Office 14500 N. Northsight Blvd. Ste. 221        Scottsdale, AZ 85260  Phone: (480) 946-2411 Fax: (480) 946-2446]

Cleaned data: [Official Policy Statement In 2013, Congressman David Schweikert and Majority Leader Kevin McCarthy teamed up to co-found the Congressional Valley Fever Task Force, bringing much needed awareness to the disease of coccidioidomycosis. The disease, more commonly known as Valley Fever, is prominent in Maricopa County as well as the rest of Arizona and southern California. The goal of the Valley Fever Task Force is to share information with stakeholders in the medical and scientific fields to foster new advancements in prevention and treatment as well as work with community organizations to help educate individuals on the disease. As Congressman Schweikert stated, “This task force is a much needed step toward raising awareness for this terrible disease and someday soon finding a cure. Valley Fever has silently affected entire communities in the southwest including our family, friends, and even beloved pets. I am hopeful that this working group will bring awareness, reduce the risk of misdiagnoses, and bring about a cure within the decade.” In the 113th Congress, Congressman Schweikert successfully led the effort to have coccidioides spp., the pathogens that cause Valley Fever, listed as qualifying pathogens under the GAIN Act of 2011. This effort granted Valley Fever the title of “orphan disease” with the Food and Drug Administration (FDA), meaning any treatments or future cures will be given priority and fast tracked through the often arduous FDA drug approval process. Valley Fever severely affects many who work outdoors, farmers, and those in the construction industry. Every branch of the United States’ military has bases in the endemic regions, putting our military men and women at risk.]

\subsection{Text Topic Labeling}
The following prompt is utilized to label the topic of the text with {\tt GPT-4o-mini} and {\tt Llama-3.1-8b}:

\textit{\textbf{Prompt 2.1:}} Identify the primary topic of the text I provide. Select one from the following list: constituent service, economy, health, foreign policy, immigration, gun control, abortion, racial and ethnic inequality, climate change, election, or other. Respond with the topic only: \{text\}

\subsection{Text stance labeling}

The following prompt is used to label the stance of the text with {\tt GPT-4o-mini} and {\tt Llama-3.1-8b}:

\textit{\textbf{Prompt 3.1:}} Evaluate the political stance revealed in the text I provide. Assign a score based on the following scale: -10 for the most liberal (left), 10 for the most conservative (right), and 0 for neutral. Respond with the score only: \{text\}

\begin{figure}[!tb]
    \centering
    \includegraphics[width= .9 \linewidth]{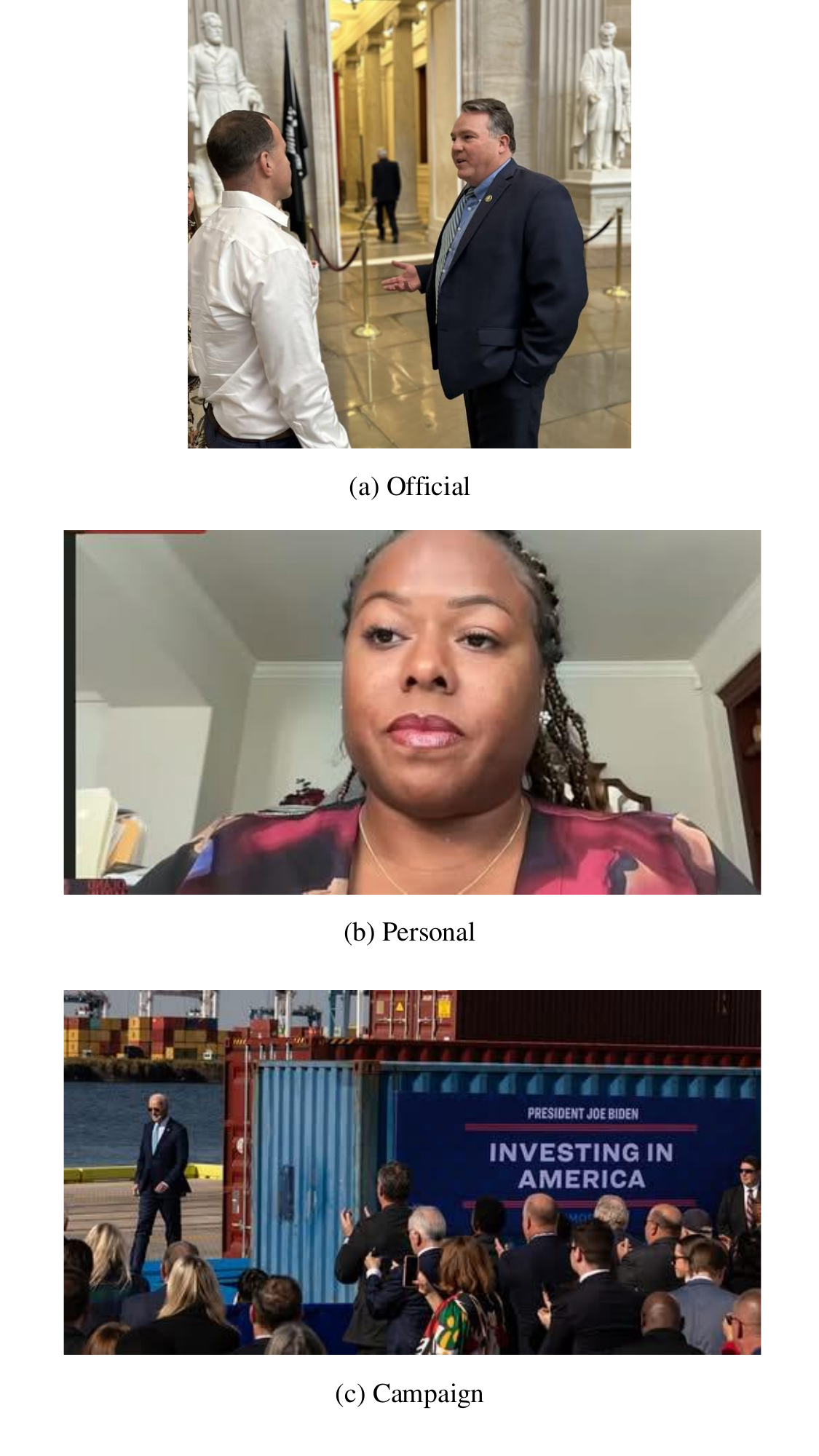}
    \caption{Examples of Cotent Labeling Generated by {\tt LLaVA-1.5-7b}}
    \label{fig: img_content}
\end{figure}

\begin{figure}[!t]
    \centering
    \includegraphics[width= .9 \linewidth]{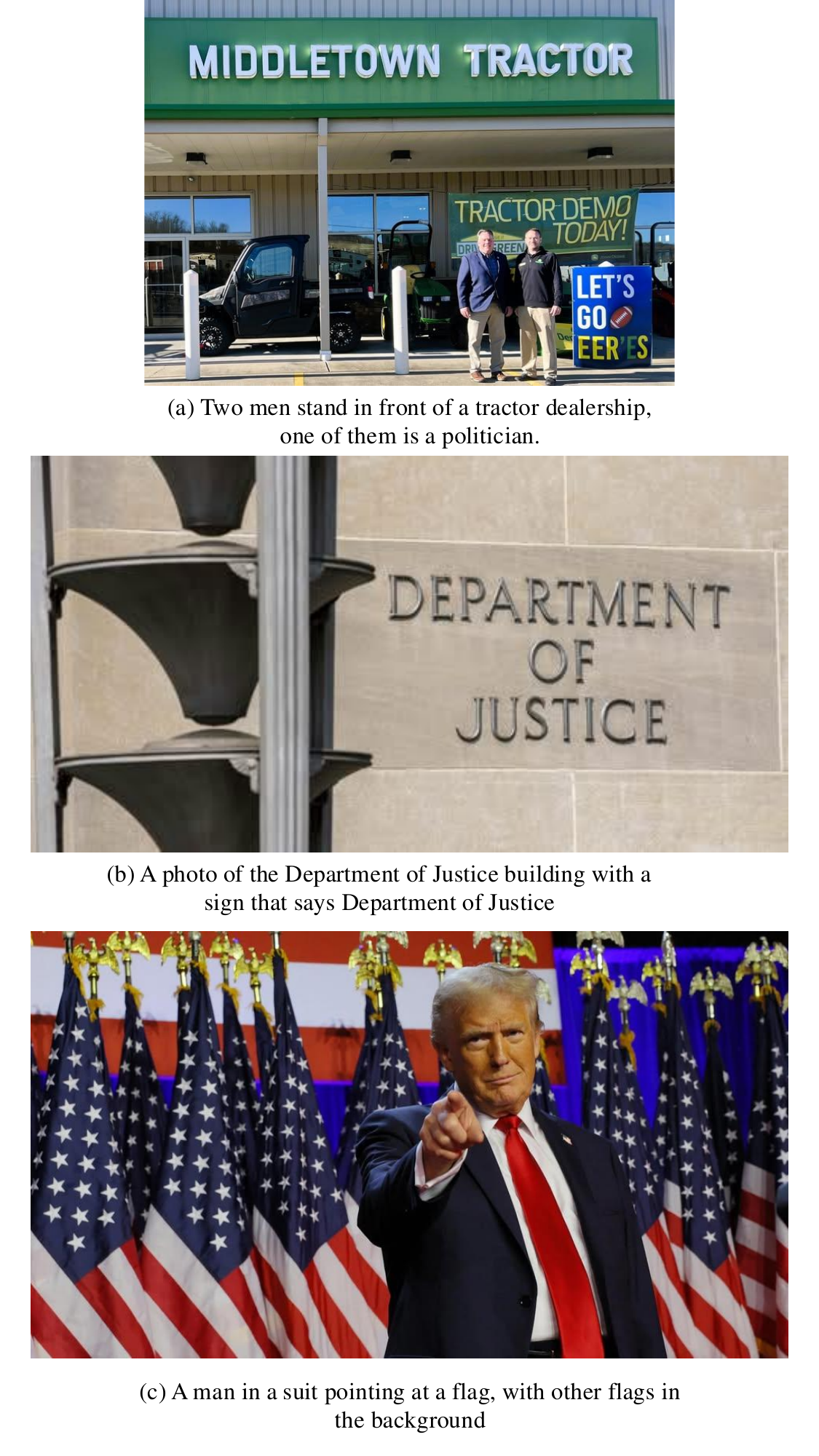}
    \caption{Examples of One-Sentence Descriptions Generated by {\tt LLaVA-1.5-7b}}
    \label{fig: img_sentence}
\end{figure}

The following prompt is utilized to label image features extracted from Facebook with {\tt LLaVA-1.5-7b}:

\textit{\textbf{Prompt 4.1:}} Classify the photo into one or more of the following categories: official, campaign, policy, or personal. Respond only with the category or categories the image belongs to.

\textit{\textbf{Example 4.1:}} Figure~\ref{fig: img_content} shows the classification examples of {\tt LLaVA-1.5-7b} on image content.

Next, we use the following prompt to {\tt LLaVA-1.5-7b} to summarize the content in the images to prepare for the further extraction of presence of people information from the data:

\textit{\textbf{Prompt 4.2:}} Analyze the photo from the official account of a U.S. legislator and provide a one-sentence description of the people shown and the overall content of the photo.

\textit{\textbf{Example 4.2:}} Figure~\ref{fig: img_sentence} presents the examples of the one-sentence summary made by {\tt LLaVA-1.5-7b}.

The following prompt is used to create the categories of the presence of people using {\tt GPT-4o-mini}:

\textit{\textbf{Prompt 4.3:}} Based on the description of the image, classify the number of people in the image into one of the following categories: individual, group, or no people. Respond only with the appropriate category: \{text\}

\subsection{Post Frequency Distribution of the Top 10 Most Active Legislators on Each Platform}

\begin{table}[H]
\centering
\resizebox{.9\linewidth}{!}{%
\begin{tabular}{lrrc}
\hline
\textbf{Congress Member} & \textbf{Frequency} & \textbf{Percentage (\%)} & \textbf{Party} \\
\hline
Jon Tester         & 80 & 1.32 & D \\
Gary Palmer        & 69 & 1.14 & R \\
Tim Walberg        & 64 & 1.06 & R \\
Marcy Kaptur       & 60 & 0.99 & D \\
Lauren Boebert     & 57 & 0.94 & R \\
Jesus Garcia       & 56 & 0.93 & D \\
Doug Lamalfa       & 51 & 0.84 & R \\
Bill Foster        & 50 & 0.83 & D \\
Maria Cantwell     & 49 & 0.81 & D \\
David Schweikert   & 48 & 0.79 & R \\
\hline
\end{tabular}%
}
\caption{Post Frequency Distribution of the Top 10 Most Active Legislators on Official Websites}
\label{tab:website_top10}
\end{table}

\begin{table}[H]
\centering
\resizebox{.9\linewidth}{!}{%
\begin{tabular}{lrrc}
\hline
\textbf{Congress Member} & \textbf{Frequency} & \textbf{Percentage (\%)} & \textbf{Party} \\
\hline
Debbie Lesko              & 467 & 0.57 & R \\
John Kennedy              & 460 & 0.56 & R \\
Charles Schumer           & 457 & 0.55 & D \\
Thomas Kean               & 456 & 0.55 & R \\
Joseph Morelle            & 455 & 0.55 & D \\
Michael Lawler            & 447 & 0.54 & R \\
Don Bacon                 & 447 & 0.54 & R \\
Debbie Wasserman Schultz  & 447 & 0.54 & D \\
Josh Gottheimer           & 446 & 0.54 & D \\
Anthony Desposito         & 444 & 0.54 & R \\
\hline
\end{tabular}%
}
\caption{Post Frequency Distribution of the Top 10 Most Active Legislators on Facebook}
\label{tab:fb_top10}
\end{table}

\begin{table}[H]
\centering
\resizebox{.9\linewidth}{!}{%
\begin{tabular}{lrrc}
\hline
\textbf{Congress Member} & \textbf{Frequency} & \textbf{Percentage (\%)} & \textbf{Party} \\
\hline
Amy Klobuchar       & 1148 & 1.00 & D \\
Mark Green          & 925  & 0.80 & R \\
Marcus Molinaro     & 797  & 0.69 & R \\
Sherrod Brown       & 714  & 0.62 & D \\
Beth Van Duyne      & 711  & 0.62 & R \\
Collin Allred       & 673  & 0.58 & D \\
Marsha Blackburn    & 653  & 0.57 & R \\
Michael Lawler      & 640  & 0.56 & R \\
Brian Mast          & 613  & 0.53 & R \\
Frank Pallone       & 610  & 0.53 & D \\
\hline
\end{tabular}%
}
\caption{Post Frequency Distribution of the Top 10 Most Active Legislators on X (Twitter)}
\label{tab:twitter_top10}
\end{table}

\begin{table}[H]

\centering
\resizebox{.9\linewidth}{!}{%
\begin{tabular}{lrrc}
\hline
\textbf{Congress Member} & \textbf{Frequency} & \textbf{Percentage (\%)} & \textbf{Party} \\
\hline
Jon Ossoff         & 303 & 10.50 & D \\
Sherrod Brown      & 303 & 10.50 & D \\
Katie Boyd Britt   & 205 & 7.10  & R \\
Collin Allred      & 203 & 7.03  & D \\
Jacky Rosen        & 191 & 6.62  & D \\
Lauren Underwood   & 125 & 4.33  & D \\
Bernard Sanders    & 119 & 4.12  & I \\
J.D. Vance         & 116 & 4.02  & R \\
Ruben Gallego      & 115 & 3.98  & D \\
Patty Murray       & 106 & 3.67  & D \\
\hline
\end{tabular}%
}
\caption{Post Frequency Distribution of the Top 10 Most Active Legislators on TikTok}
\label{tab:tiktok_top10}
\end{table}

\subsection{Official vs. Non-official Twitter Accounts}

\begin{figure}[H]
    \centering
    \includegraphics[width=.45\textwidth]{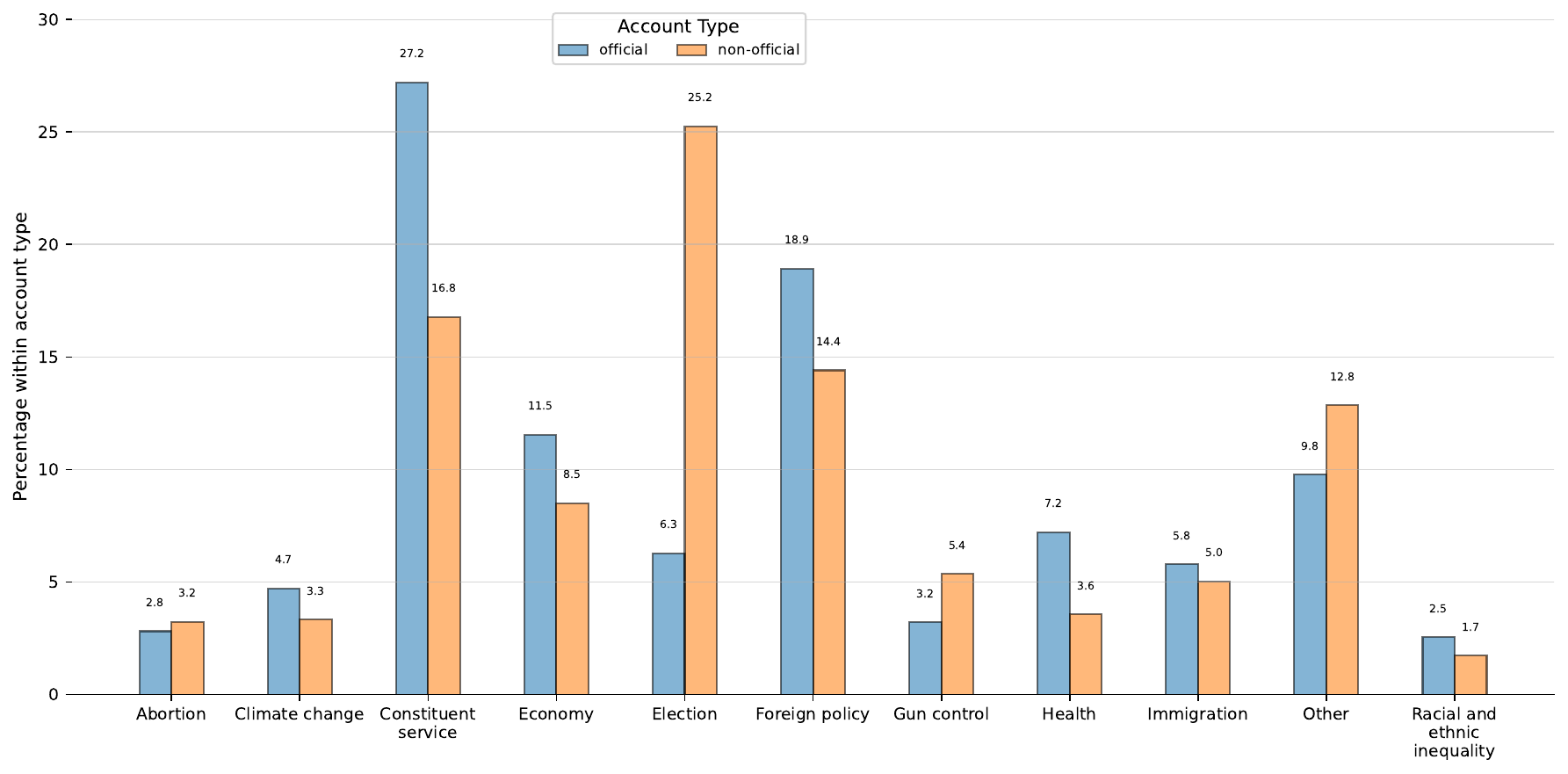}
    \caption{Topic Distribution of Official and Non-official X (Twitter) Account}
    \label{fig: balance_topic}
\end{figure}

\begin{figure}[H]
  \centering
  \includegraphics[width=.45\textwidth]{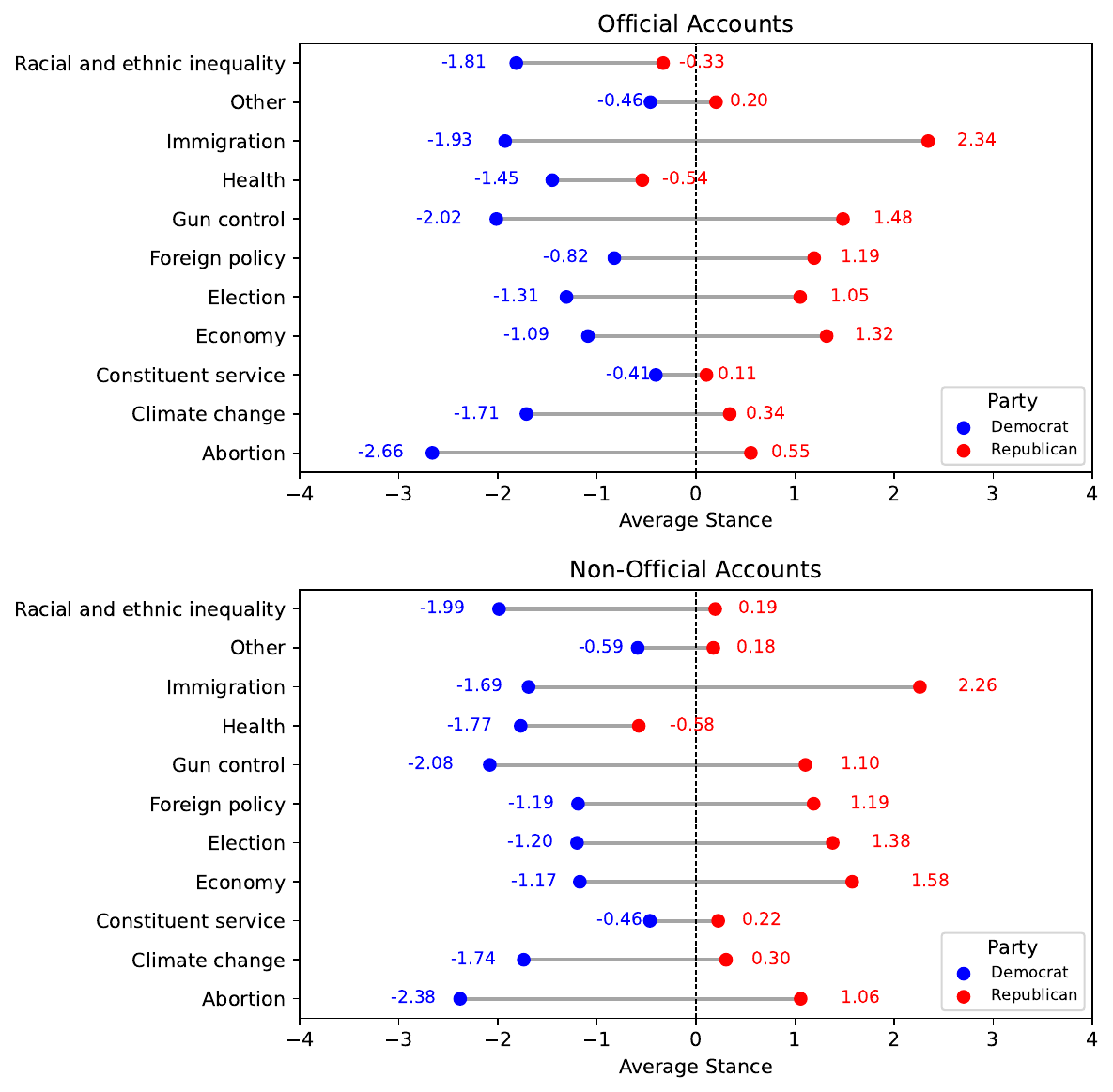}
  \caption{Stance Divergence Across Topics by Partisanship, Comparing Official and Non-Official X (Twitter) Accounts}
  \label{fig: balance_stance}
\end{figure}

\begin{figure}[H]
  \centering
  \includegraphics[width= .9 \linewidth]{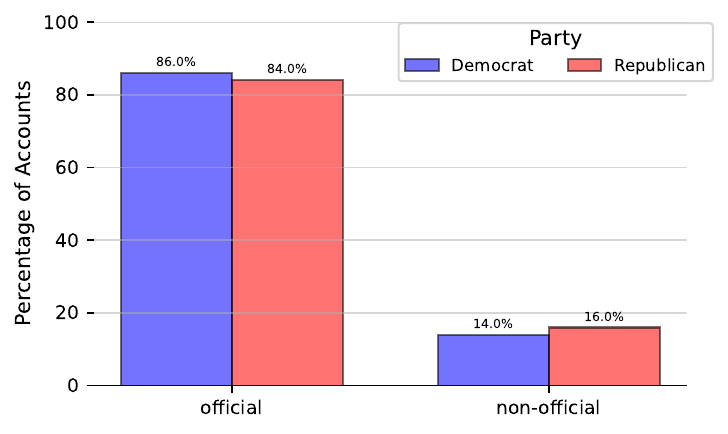}
  \caption{Partisanship Distribution of Official and Non-Official X (Twitter) Accounts}
  \label{fig: balance_party}
\end{figure}

\end{document}